\documentclass[twocolumn,secnumarabic,amssymb,amsmath,superscriptaddress,nofootinbib,tightenlines,nobibnotes,aps,prc]{revtex4-2}

\usepackage{graphicx,xcolor}
\usepackage[unicode]{hyperref}

\begin{document}


\title{Superscaling analysis of inclusive electron and (anti)neutrino scattering within the coherent density fluctuation model}

\author{M.V.~Ivanov\footnote{martin.inrne@gmail.com}}
\affiliation{Institute for Nuclear Research and Nuclear Energy, Bulgarian Academy of Sciences, Sofia 1784, Bulgaria}
\author{A.N.~Antonov}
\affiliation{Institute for Nuclear Research and Nuclear Energy, Bulgarian Academy of Sciences, Sofia 1784, Bulgaria}

\date{\today}

\begin{abstract}
The experimental data from quasielastic electron and (anti)neutrino scattering on $^{12}$C are reanalyzed in terms of a new scaling variable $\psi^*$ suggested by the interacting relativistic Fermi gas with scalar and vector interactions, which is known to generate a relativistic effective mass for the interacting nucleons. We construct a new scaling function $f^\text{QE}(\psi^*)$ for the inclusive lepton scattering from nuclei within the coherent density fluctuation model (CDFM). The latter is a natural extension of the relativistic Fermi gas model to finite nuclei. In this work, on the basis of the scaling function obtained within CDFM with a relativistic effective mass $m_N^* =0.8 m_N$, we calculate and compare the theoretical predictions with a large set of experimental data for inclusive ($e,e'$) and (anti)neutrino cross sections. The model also includes the contribution of weak two-body currents in the two-particle two-hole sector, evaluated within a fully relativistic Fermi gas. Good agreement with experimental data is found over the whole range of electron and (anti)neutrino energies.
\end{abstract}

\maketitle

\section{Introduction\label{sec:introduction}}

The superscaling phenomenon was firstly considered within the framework of the Relativistic Fermi Gas (RFG) model~\cite{PhysRevC.38.1801, Barbaro1998137, PhysRevLett.82.3212, PhysRevC.60.065502, PhysRevC.65.025502, BCD+04}, where a properly defined function of the scaling $\psi$-variable was introduced. At large transferred momentum $q=|\mathbf{q}|$ ($q>500$~MeV/c) the latter does not depend on $q$ and the mass number. As pointed out in~\cite{PhysRevC.60.065502}, however, the actual nuclear dynamical content of the superscaling is more complex than that provided by the RFG model. It was observed that the experimental data have a superscaling behavior in the low-$\omega$ side ($\omega$ being the transfer energy) of the quasielastic peak for large negative values of $\psi$ (up to $\psi\approx -2$), while the predictions of the RFG model are $f(\psi)=0$ for $\psi \leq -1$. This imposes the consideration of the superscaling in realistic finite systems. One of the approaches to do this was developed~\cite{A7, A8} in the CDFM~\cite{anton1, anton2, antonov_bjp_1979, antonov_nucleon_1980, antonov_spectral_1982, antonov_extreme_1985, antonov_natural_1989, PhysRevC.50.164} which is related to the $\delta$-function limit of the generator coordinate method \cite{A7, PhysRev.108.311}. It was shown in~\cite{A7, A8, A9} that the superscaling in nuclei can be explained quantitatively on the basis of the similar behavior of the high-momentum components of the nucleon momentum distribution in light, medium and heavy nuclei. It is well known that the latter is related to the effects of the \emph{NN} correlations in nuclei (see, \emph{e.g.}~\cite{anton1, anton2}).

In our previous works~\cite{A7, A8, A9, A10} we obtained the CDFM scaling function $f(\psi)$ starting from the RFG model scaling function $f_\text{RFG}(\psi)$ and convoluting it with the weight function $|F(x)|^2$ that is related equivalently to either the density $\rho(r)$ or the nucleon momentum distribution $n(k)$ in nuclei. Thus, the CDFM scaling function is an infinite superposition of weighted RFG scaling functions. This approach improves upon RFG and enables one to describe the scaling function for realistic finite nuclear systems. The CDFM scaling function has been used to predict cross sections for several processes such as the inclusive electron scattering in the QE and $\Delta$- regions~\cite{A10, A13} and neutrino (antineutrino) scattering both for charge-changing (CC)~\cite{A13} and for neutral-current (NC)~\cite{A12} processes. In our work~\cite{A10} we reproduce experimental data of the inclusive electron scattering in the QE-region using CDFM scaling function which is obtained by the parameterizing the RFG scaling function and by the coefficient $c_1$, which helps us to account for the experimental fact of the asymmetry of the scaling function. The value of the coefficient $c_1$ ($c_1 \neq 3/4$) is taken in accordance with the empirical data ($c_1$ depends on the value of the momentum transfer in the QE peak).

In the present work we follow Ref.~\cite{Ama15}, where the $\psi^*$ scaling idea is explored in the context of the Relativistic Mean Field (RMF) for nuclear matter. The new scaling function $f^*(\psi^*)$ including dynamical relativistic effects~\cite{Ama15, Ama17, Mar17, Ama18} is introduced through an effective mass into its definition. The resulting superscaling approach with relativistic effective mass (SuSAM*) model describes a large amount of the electron data lying inside a phenomenological quasielastic band, and it has been extended recently successfully to the neutrino and antineutrino sector~\cite{Rui18} giving a fair agreement with the data. An enhancement of the SuSAM* model is detailed in Refs.~\cite{PhysRevC.104.025501, PhysRevD.104.113006, PhysRevD.108.013007, PhysRevD.108.113006}, where the responses of 2p2h meson exchange currents (MEC) are calculated consistently with the mean field model in nuclear matter. This is achieved by incorporating an effective mass and vector energy for the nucleon, thereby explicitly including the same medium modifications as the quasielastic responses. A systematic review of experimental data on quasielastic neutrino scattering reveals a reasonable agreement with the theoretical predictions derived from the extended SuSAM* model~\cite{PhysRevD.108.113006}.

The SuSAM* model was first developed using the set of $^{12}$C data~\cite{Ama15, Ama17} and later applied to other nuclei in~\cite{Mar17}. In Ref.~\cite{Ama15} was obtained the best value of the effective mass $M^* = {m_N^*}/{m_N} = 0.8$, which we use in our present consideration. This value provides the best scaling behavior of the data with a large fraction of data concentrated around the universal scaling function of the relativistic Fermi gas
\begin{equation}
f_{\rm RFG}(\psi^*) = \frac{3}{4}(1-\psi^*{}^2) \theta(1-\psi^*{}^2).
\label{rfg*}
\end{equation}
The $\psi^*$ variable [see Eqs.~(\ref{eps0*})--(\ref{psi*}) given below] was inspired by the mean-field theory, that provides a reasonable description of the quasielastic response function~\cite{Ros80, Ser86}. The important point is that in the interacting RFG model the vector and scalar potentials generate an effective mass $m^*_N$ for the nucleon in the medium.

Our present approach, called CDFM$_{M^*}$ (CDFM with $M^*$), uses scaling function obtained within the CDFM model. It keeps the gauge invariance and describes the dynamical enhancement of both the lower components of the relativistic spinors and the transverse response function. Here we mention that the SuperScaling Analyses (SuSA) violates the gauge invariance because it introduces an energy shift to account for separation energy.

The paper is organized as follows: In Section~\ref{sec:scheme} we review in brief the theoretical scheme for obtaining the CDFM$_{M^*}$ scaling function and the general formalism for the description of the $(e,e')$ and (anti)neutrino CC quasielastic double differential cross sections. In Section~\ref{sec:results} we show our main results for the inclusive $(e,e')$ and (anti)neutrino CC quasielastic double differential cross sections, and finally, in Section~\ref{sec:conclusions} we draw our conclusions and outline our future plans or prospects related to the conclusions of the present work.

\section{Theoretical scheme\label{sec:scheme}}

The model developed in Refs.~\cite{Ama15, Ama17, Mar17, Ama18} is inspired by the RMF in nuclear matter. In this model the single-nucleon excitations with initial nucleon energy $E=\sqrt{{\bf p}^2+m_N^*{}^2}$ in the mean field is considered. The final momentum of the nucleon is ${\bf p}'={\bf p}+{\bf q}$ and its energy is $E'=\sqrt{{\bf p}'{}^2+m_N^*{}^2}$. Note that initial and final nucleons have the same effective mass $m_N^*$.

In the case of the $(e,e')$ scattering the quasielastic cross section is written in the Rosenbluth form
\begin{equation}
\frac{d\sigma}{d\Omega'd\epsilon'} = \sigma_{\rm Mott}\left( v_L R_L + v_T R_T \right),\label{xsecel}
\end{equation}
where $\sigma_{\rm Mott}$ is the Mott cross section, $v_L= Q^4/q^4$ and $v_T=\tan^2(\theta/2)-Q^2/2q^2$, with $\theta$ the scattering
angle, the energy transfer $\omega$ and $3$-momentum transfer ${\bf q}$. The four-momentum transfer is $Q^2 = \omega^2 - q^2 < 0$. The nuclear response functions can be written in the factorized form for $K=L,T$
\begin{equation}
R_K = r_K f^*(\psi^*), \label{factorization}
\end{equation}
where $r_L$ and $r_T$ are the single-nucleon contribution, taking into account the Fermi motion
\begin{equation}
r_K = \frac{\xi_F^*}{m^*_N {\eta_F^*}^3 \kappa^*} (Z U^p_K+NU^n_K).
\label{single}
\end{equation}
In our calculations we use the scaling function $f^*(\psi^*)$ obtained within RFG model [Eq.~(\ref{rfg*})] as well as the one within the CDFM$_{M^*}$ model [see Eq.~(\ref{eq:8}) given below]. They depend only on the new scaling variable $\psi^*$, that is the minimum kinetic energy of the initial nucleon divided by the kinetic Fermi energy. The minimum energy allowed for a nucleon inside the nucleus to absorb the virtual photon (in units of $m_N^*$) is
\begin{equation}
\epsilon_0^*={\rm Max}
\left\{
       \kappa^*\sqrt{1+\frac{1}{\tau^*}}-\lambda^*, \epsilon_F^*-2\lambda^*\label{eps0*}
\right\},
\end{equation}
where the dimensionless variables are
\begin{eqnarray*}
\lambda^*  &=& \omega/(2m_N^*), ~
\kappa^*   =  q/(2m_N^*),~
\tau^*  =  {\kappa^*}^2-{\lambda^*}^2, \\
\eta_F^* & = &  k_F/m_N^*,~
\xi_F^*  =  \sqrt{1+{\eta_F^*}^2}-1,~
\epsilon_F^* = \sqrt{1+{\eta_F^*}^2}.
\end{eqnarray*}
The definition of the scaling variable is given
\begin{equation}
\psi^* = \sqrt{\frac{\epsilon_0^*-1}{\epsilon_F^*-1}} {\rm sgn} (\lambda^*-\tau^*),\label{psi*}
\end{equation}
where $\psi^*$ is negative to the left of the quasielastic peak ($\lambda^* < \tau^*$) and positive on the right side.

The single nucleon response functions are
\begin{eqnarray}
U_L &=& \frac{{\kappa^*}^2}{\tau^*}
\left[ (G^*_E)^2 + \frac{(G_E^*)^2 + \tau^* (G_M^*)^2}{1+\tau^*}\Delta \right]
\\
U_T &=& 2\tau^*  (G_M^*)^2 + \frac{(G_E^*)^2 + \tau^* (G_M^*) ^2}{1+\tau^*}\Delta
\end{eqnarray}
where the quantity $\Delta$ has been introduced
\begin{equation}
\Delta= \frac{\tau^*}{{\kappa^*}^2}\xi_F^*(1-\psi^*{}^2)
\left[ \kappa^*\sqrt{1+\frac{1}{\tau^*}}+\frac{\xi_F^*}{3}(1-\psi^*{}^2)\right].
\end{equation}
The electric and magnetic form factors are modified in the medium due to the effective mass according to
\begin{eqnarray}
G_E^*  =  F_1-\tau^* \frac{m^*_N}{m_N} F_2,\quad
G_M^*  = F_1+\frac{m_N^*}{m_N} F_2.  \label{GE&GM}
\end{eqnarray}
For the free Dirac and Pauli form factors, $F_1$ and $F_2$, we use the Galster parametrization~\cite{GALSTER1971221}.

The scaling function in the CDFM was obtained starting from that in the RFG model~\cite{PhysRevC.38.1801, Barbaro1998137, PhysRevLett.82.3212, PhysRevC.60.065502} in two equivalent ways: on the basis of the local density distribution $\rho(r)$ and of the nucleon momentum distribution $n(k)$. This allows one to study simultaneously the role of the \emph{NN} correlations included in $\rho(r)$ and $n(k)$ in the case of the superscaling phenomenon. To explore these properties the scaling function has been derived in two ways in CDFM in~\cite{A8}. When using the density distribution $\rho(r)$ the scaling function is:
\begin{equation}
f^\text{QE}(\psi^*)= \int_{0}\limits^{\infty}  |F(x)|^{2} \,f_\text{RFG}^\text{QE}[\psi^*(x)]\,{\mathop{}\!\mathrm{d}} x, \label{eq:1}
\end{equation}
with a weight function of the form
\begin{equation}
|F(x)|^{2}=-\frac{1}{\rho_{0}(x)} \left. \frac{d\rho(r)}{dr}\right |_{r=x}, \label{eq:2}
\end{equation}
where
\begin{equation}
\rho_{0}(x)=\frac{3A}{4\pi x^{3}}.\label{eq:3}
\end{equation}
$f_\text{RFG}^\text{QE}[\psi^*(x)]$ with $\psi^*(x)=\dfrac{k_{F}x\psi^*}{\alpha}$ is the scaling function related to the RFG model [see Eq.~(\ref{rfg*})]
\begin{equation}
f_\text{RFG}^\text{QE}[\psi^*(x)] = \displaystyle \frac{3}{4} \left[\! 1\!-\!\left( \frac{k_F x \psi^*}{\alpha} \right)^{2}\!\right]
\theta \left(\!1-\!\left(\dfrac{k_F x\psi^*}{\alpha}\right)^2\!\right), \label{eq:4}
\end{equation}
where $\alpha=(9\pi A/8)^{1/3}\simeq 1.52A^{1/3}$ and therefore
\begin{equation}
f^\text{QE}(\psi^*)= \int_{0}\limits^{\alpha/(k_{F}|\psi^*|)}  |F(x)|^{2} \,f_\text{RFG}^\text{QE}[\psi^*(x)]\,{\mathop{}\!\mathrm{d}} x. \label{eq:5}
\end{equation}
In Eqs.~(\ref{eq:1}) and (\ref{eq:4}) the Fermi momentum $k_{F}$ is not a free parameter for different nuclei as it is in the RFG model, but $k_{F}$ is calculated within the CDFM (or CDFM$_{M^*}$) for each nucleus using the corresponding expressions:
\begin{multline}
k_F= \int\limits_{0}^{\infty} |F(x)|^2 \,k_{x}(x)\, {\mathop{}\!\mathrm{d}} x=  \int\limits_{0}^{\infty}  |F(x)|^{2}\,\frac{\alpha}{x}\, {\mathop{}\!\mathrm{d}} x=\\= \frac{4\pi(9\pi)^{1/3}}{3A^{2/3}} \int\limits_{0}^{\infty} \rho(r)\, r\, {\mathop{}\!\mathrm{d}} r \label{eq:6}
\end{multline}
when the condition
\begin{equation}
\lim_{x\rightarrow \infty} \left[ \rho(r)\,r^2 \right]=0 \label{eq:7}
\end{equation}
is fulfilled. As shown in~\cite{A8}, the integration in Eq.~(\ref{eq:5}), using Eq.~(\ref{eq:2}), leads to the explicit relationship of the scaling function with the density distributions:
\begin{multline}
f^\text{QE}(\psi^*)= \frac{4\pi}{A}\int\limits_{0}^{\alpha/(k_{F}|\psi^*|)} \rho(x) \Bigg[ x^2 f_\text{RFG}^\text{QE}[\psi^*(x)]+\\+ \frac{x^3}{3} \frac{\partial f_\text{RFG}^\text{QE}[\psi^*(x)]}{\partial x} \Bigg]\, {\mathop{}\!\mathrm{d}} x. \label{eq:8}
\end{multline}
We also note that in the consideration up to here the CDFM$_{M^*}$ model scaling function $f^\text{QE}(\psi^*)$ is symmetric under the change of $\psi^*$ by $-\psi^*$. 

\begin{figure}[tb]
\centering\includegraphics[width=1\linewidth]{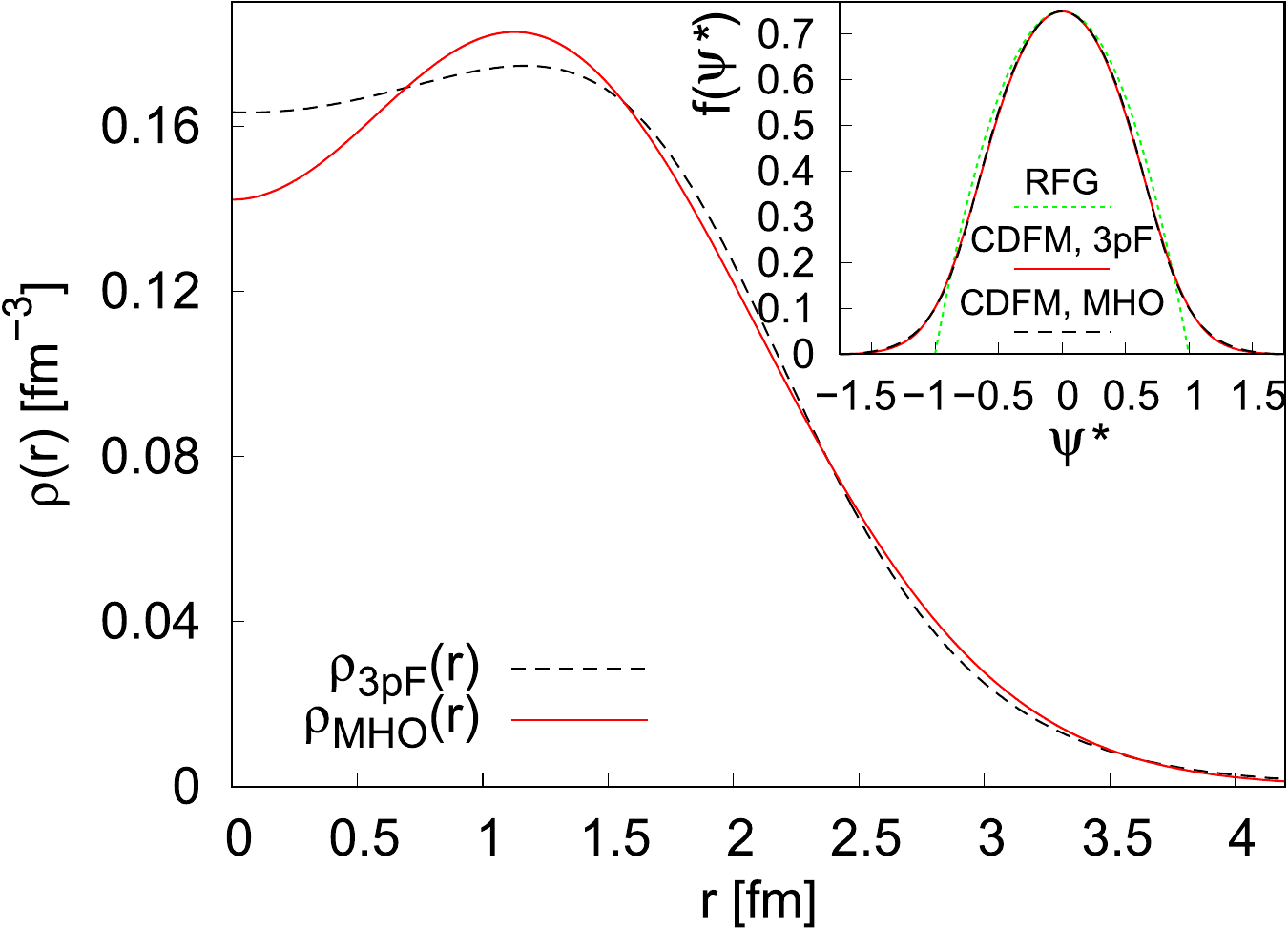}
\caption{(Color online) The nucleon density distribution of the $^{12}$C nucleus using two empirical distributions of the protons~\cite{Patt2003}: Modified Harmonic Oscillator (\emph{MHO}) and three parameter Fermi (\emph{3pF}) parameterizations. As inset depicts the corresponding scaling functions obtained within CDFM$_{M^*}$ [using Eqs.~(\ref{eq:6}) and (\ref{eq:8})] and RFG$_{M^*}$ [Eq.~(\ref{rfg*})] models are given.\label{fig:fpsi}}
\end{figure}

In this work we use the empirical density distribution of protons to determine the weight function $|F(x)|^{2}$ [see Eqs.~(\ref{eq:1}) and (\ref{eq:2})] and the corresponding scaling function in the QE-region $f^\text{QE}(\psi^*)$ within CDFM$_{M^*}$ [see Eq.~(\ref{eq:8})]. The empirical distribution of the proton parameters used in this work are those of Ref.~\cite{Patt2003}. In the case of $^{12}$C nucleus there are two parameterizations: the Modified Harmonic Oscillator (\emph{MHO}) and the three parameter Fermi (\emph{3pF}) form. In both cases (\emph{MHO} and \emph{3pF}) the condition of Eq.~(\ref{eq:7}) is fulfilled. Also, we assume that for $^{12}$C nucleus $\rho_n(r) = \rho_p(r)$. In Fig.~\ref{fig:fpsi} are presented the density distributions using \emph{MHO} and \emph{3pF} parameterizations and as inset depicts the corresponding CDFM$_{M^*}$ scaling functions in comparison with the RFG$_{M^*}$ scaling function. As can be seen, the differences between the two scaling functions obtained within CDFM$_{M^*}$ [using Eqs.~(\ref{eq:6}) and (\ref{eq:8})] are almost negligible. So that in the following results presented in the paper we use only \emph{3pF} density parametrization for calculations of the scaling function.

\begin{figure*}[tb]
\centering\includegraphics[width=.9\textwidth]{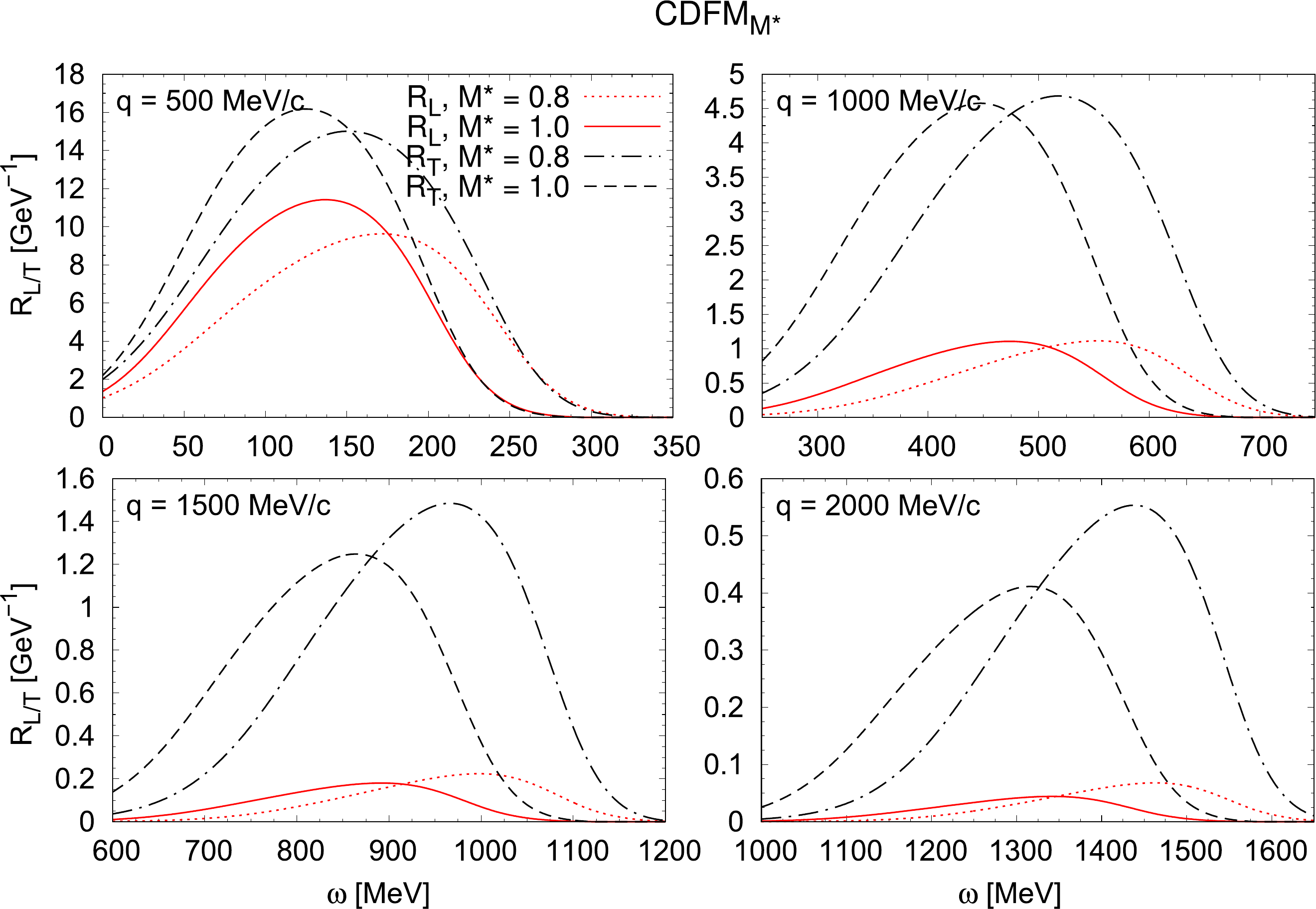}
\caption{(Color online) Longitudinal and transverse response functions of $^{12}$C in the CDFM$_{M^*}$ model, for several values of the momentum transfer. The CDFM$_{M^*}$ results for effective mass $M^* = 1$ and $0.8$ are also shown.\label{fig:RL&RT}}
\end{figure*}

In Fig.~\ref{fig:RL&RT} the longitudinal and transverse response functions for several values of the momentum transfer are shown. The CDFM$_{M^*}$ results with $M^* = 0.8$ are compared to the CDFM ones (\emph{i.e.} CDFM$_{M^*}$ with effective mass $M^* = 1$). As was shown in Ref.~\cite{Ama17}, the effect of the effective mass is a shift of the responses to higher energies, because the position of the quasielastic peak is given by $\omega \simeq \sqrt{q^2+{m_N^*}^2}-{m_N^*}$. Note that this shift gives the correct position of the quasielastic peak without need of introducing a separation energy parameter. In comparing of the ratio between $R_T$ and $R_L$ one can also observe an enhancement for the case $M^* = 0.8$, especially at larger values of the momentum transfer $q$. This is related with the known enhancement of the lower components in the relativistic spinor in the nuclear medium.

Here, it is important to note that in the present work the weight and the scaling functions which are obtained within the CDFM$_{M^*}$ model and are used in our calculations are normalized as follows:
\begin{equation}\label{norm}
  \int\limits_{0}^{\infty} |F(x)|^2 {\mathop{}\!\mathrm{d}} x =1,\quad
  \int\limits_{-\infty}^{\infty}f^\text{QE}(\psi^*){\mathop{}\!\mathrm{d}} \psi^* = 1.
\end{equation}

In the next section, we present results for the $(\nu_\mu,\mu^-)$ cross section calculated within the CDFM$_{M^*}$ model. The energies of the incident neutrino and detected muon are $\epsilon=E_\nu$, $\epsilon'=m_\mu+T_\mu$, and their momenta are ${\bf k},{\bf k}'$. The four-momentum transfer is $k^\mu-k'{}^\mu=(\omega,{\bf q})$, with $Q^2=q^2-\omega^2 > 0$. If the (anti)neutrino scattering angle is $\theta_\mu$, the double-differential cross section can be written as~\cite{Ama05a, Ama05b}
\begin{multline}
\frac{d^2\sigma}{dT_\mu d\cos\theta_\mu}
=
\sigma_0
\big\{
V_{CC} R_{CC} + 2{V}_{CL} R_{CL} + {V}_{LL} R_{LL} +\\+ {V}_{T} R_{T} \pm 2{V}_{T'} R_{T'}
\big\} \, ,\label{nuxsec}
\end{multline}
where
\begin{equation}
\sigma_0=
\frac{G^2\cos^2\theta_c}{4\pi}
\frac{k'}{\epsilon}v_0.
\end{equation}
Here $G=1.166\times 10^{-11}\quad\rm MeV^{-2} \sim 10^{-5}/ m_p^2$ is the Fermi constant, $\theta_c$ is the Cabibbo angle,
$\cos\theta_c=0.975$, and the kinematic factor $v_0= (\epsilon+\epsilon')^2-q^2$. The nuclear structure is implicitly written as  a linear combination of five nuclear response functions,  $R_K(q,\omega)$, where the fifth response function $R_{T'}$ is added (+) for neutrinos and subtracted ($-$) for antineutrinos). The $V_K$ coefficients depend only on the lepton kinematics and are independent on the details of the nuclear target.

The resulting nuclear response function $R_K$ is proportional to the single-nucleon response function $U_K$ times the scaling function $f^\text{QE}(\psi^*)$
\begin{equation} \label{cdfmm*}
R_K = \frac{{\cal N} \xi^*_F}{m^*_N \eta_F^{*3} \kappa^*}  U_K  f^\text{QE}(\psi^*).
\end{equation}
In this work we use the CDFM$_{M^*}$ scaling function obtained by Eq.~(\ref{eq:8}). It is tested in Section~\ref{sec:results} in the calculations of the inclusive $(e,e')$ quasielastic data of $^{12}$C. The kinematic coefficients $V_K$ and the single-nucleon response function $U_K$ are given in Ref~\cite{Rui18}.

\begin{figure*}[p!]
\centering\includegraphics[width=1.\textwidth]{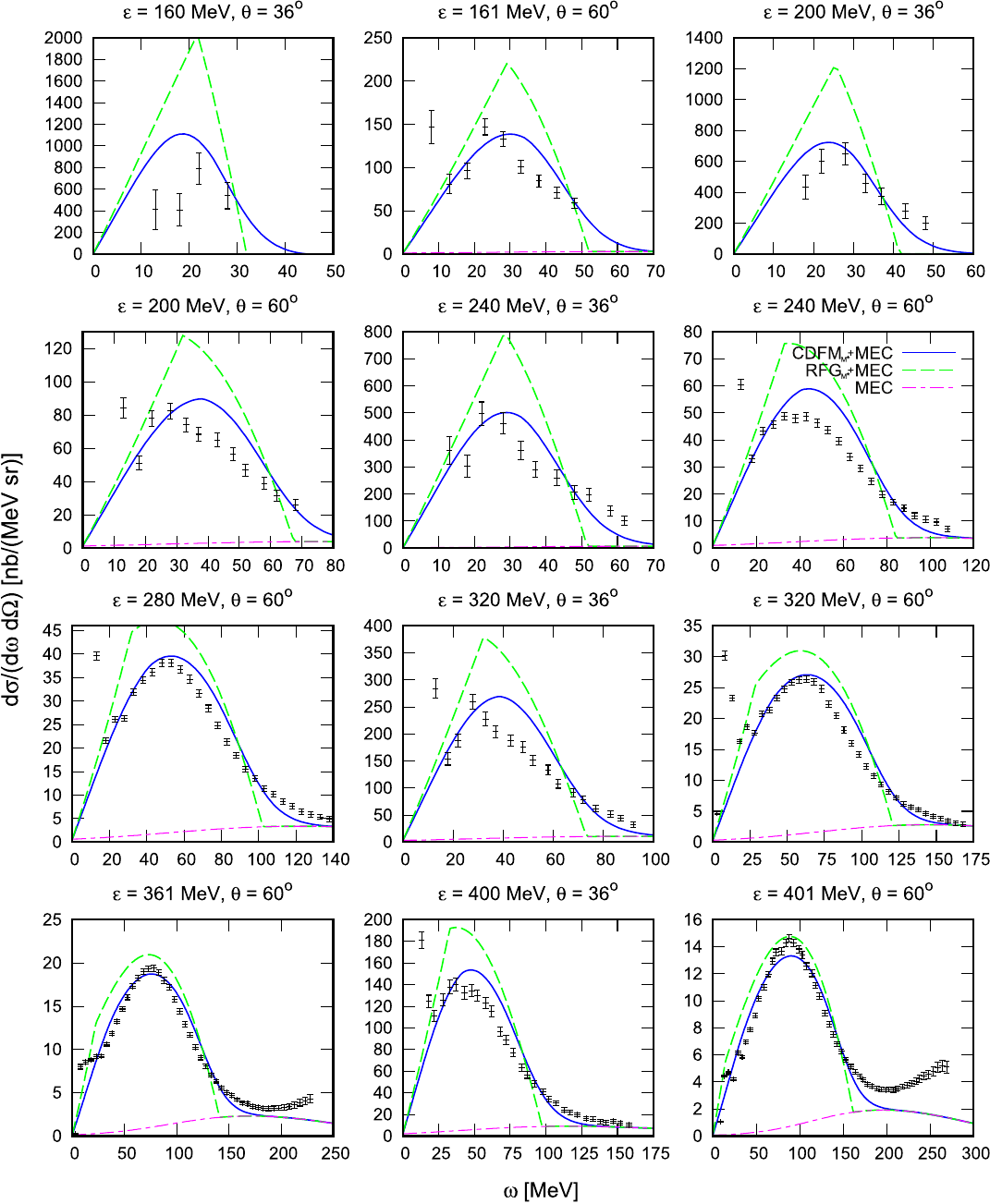}
\caption{(Color online) The CDFM$_{M^*}$ results for the inclusive $(e, e')$ cross section for several kinematics compared to the RFG$_{M^*}$ model and experimental data.\label{fig:CS1}}
\end{figure*}
\begin{figure*}[p!]
\centering\includegraphics[width=1.\textwidth]{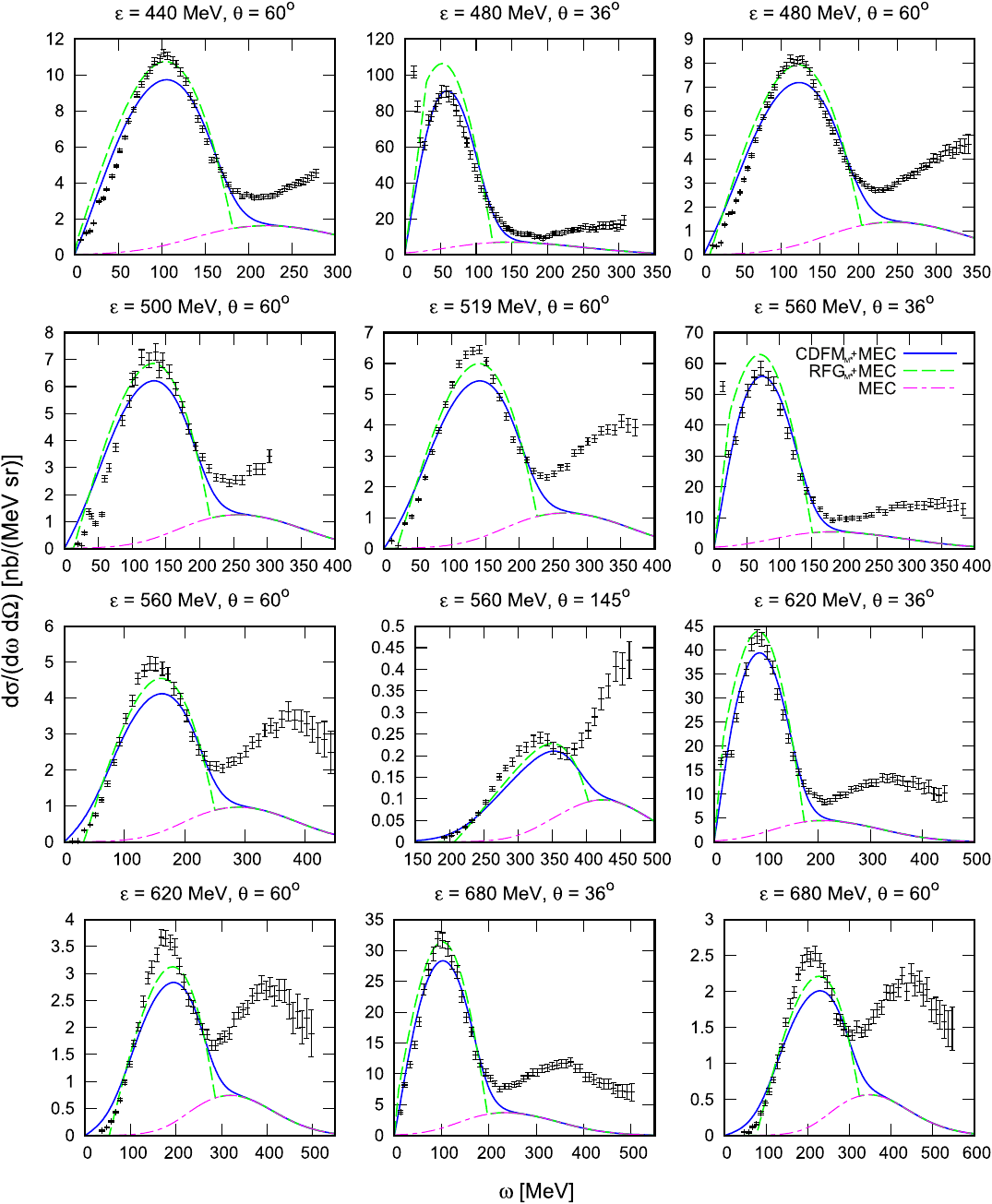}
\caption{(Color online) The CDFM$_{M^*}$ results for the inclusive $(e, e')$ cross section for several kinematics compared to the RFG$_{M^*}$ model and experimental data.\label{fig:CS2}}
\end{figure*}
\begin{figure*}[p!]
\centering\includegraphics[width=1.\textwidth]{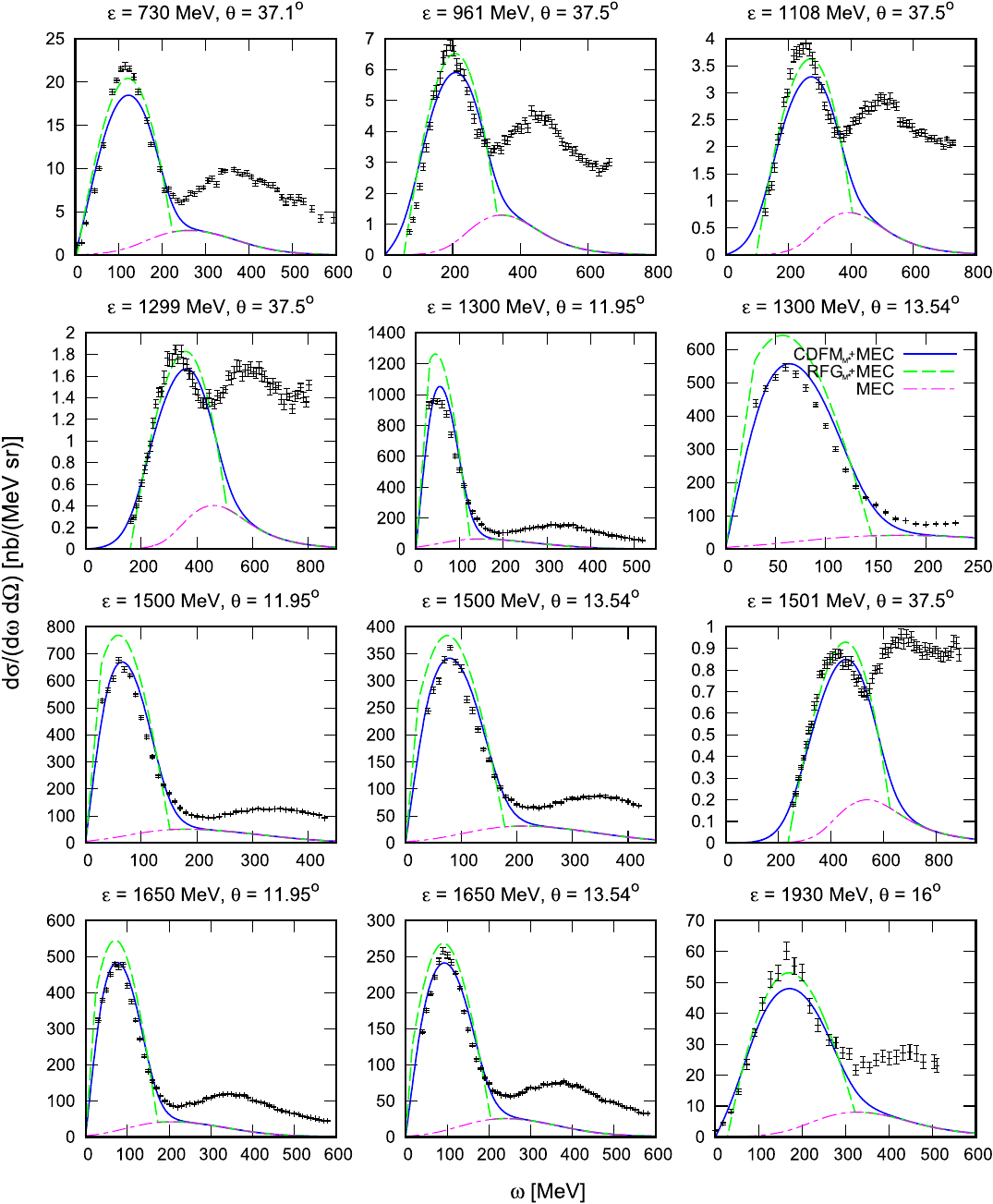}
\caption{(Color online) The CDFM$_{M^*}$ results for the inclusive $(e, e')$ cross section for several kinematics compared to the RFG$_{M^*}$ model and experimental data.\label{fig:CS3}}
\end{figure*}
\begin{figure*}[p!]
\centering\includegraphics[width=1.\textwidth]{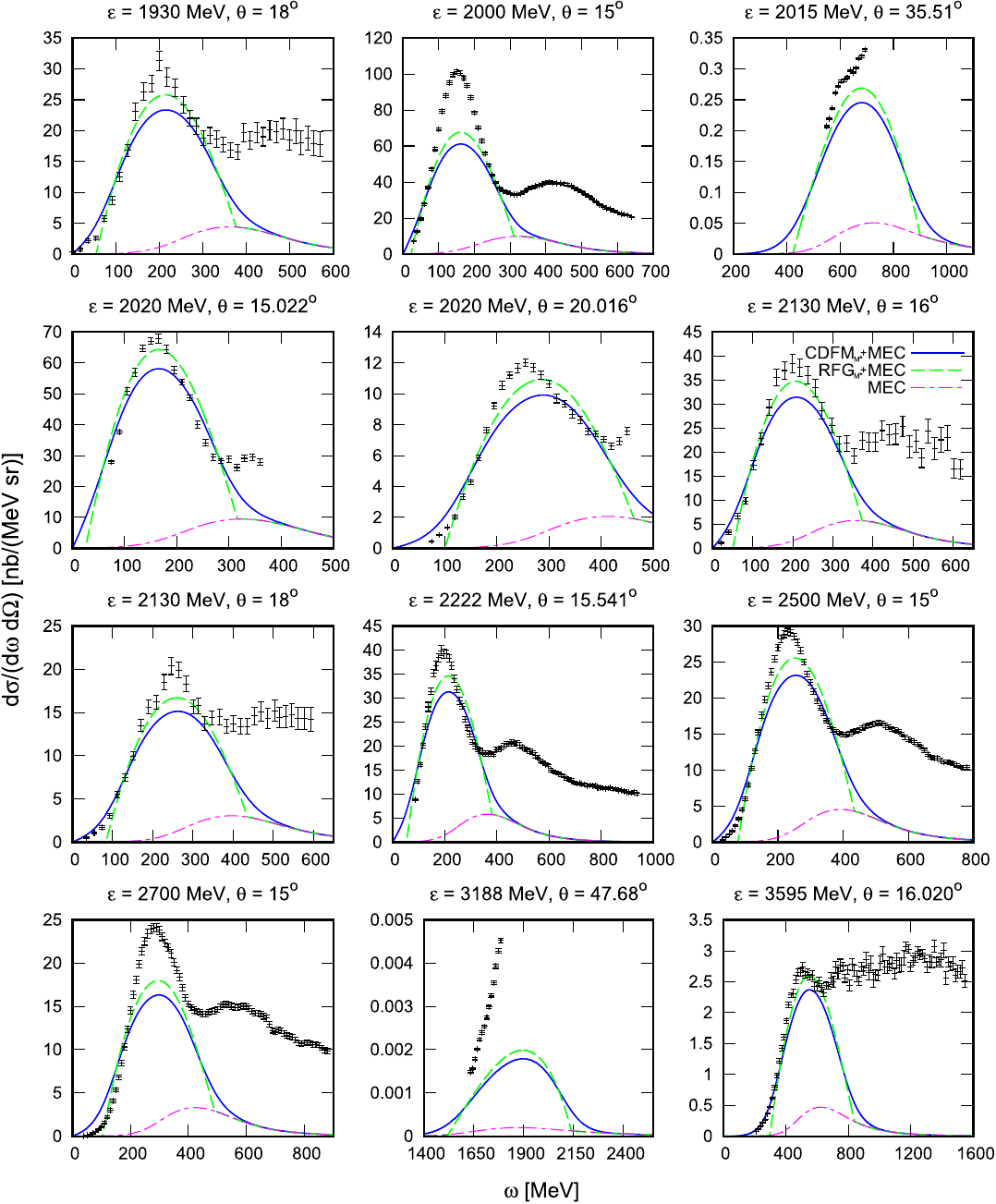}
\caption{(Color online) The CDFM$_{M^*}$ results for the inclusive $(e, e')$ cross section for several kinematics compared to the RFG$_{M^*}$ model and experimental data.\label{fig:CS4}}
\end{figure*}
\begin{figure*}[t!]
\centering\includegraphics[width=1.\textwidth]{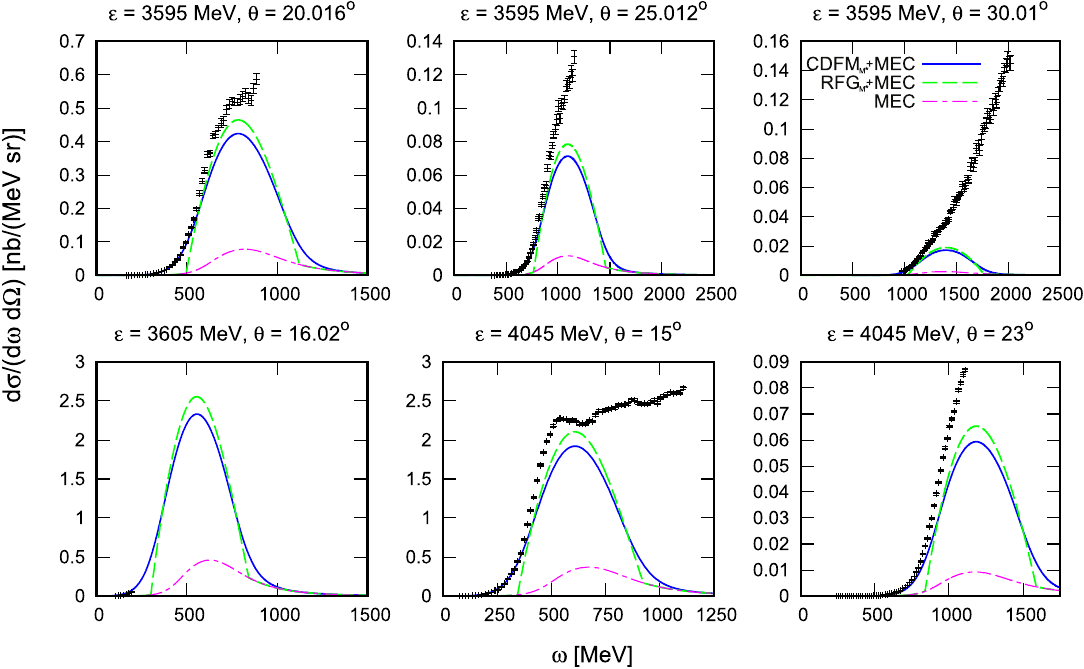}
\caption{(Color online) The CDFM$_{M^*}$ results for the inclusive $(e, e')$ cross section for several kinematics compared to the RFG$_{M^*}$ model and experimental data.\label{fig:CS5}}
\end{figure*}

\begin{figure*}[t!]
\centering\includegraphics[width=1.\textwidth]{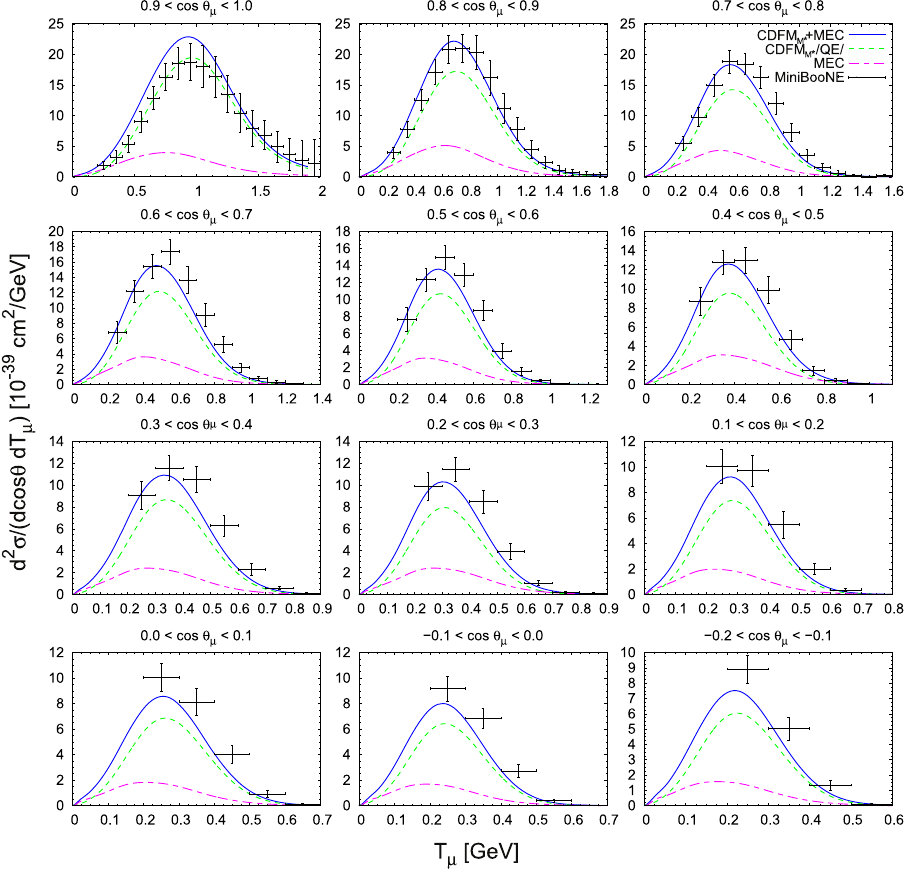}
\caption{(Color online) MiniBooNE flux-folded double differential cross section per target neutron for the $\nu_\mu$ CCQE process on $^{12}$C displayed versus the $\mu^{-}$ kinetic energy $T_\mu$ for various bins of $\cos \theta_\mu$ obtained within the CDFM$_{M^*}$ model including MEC. 2p--2h MEC and QE results are shown separately. The data are from~\cite{PhysRevD.81.092005}.\label{fig:nuMiniBooNE1}}
\end{figure*}

\begin{figure*}[t!]
\centering\includegraphics[width=1.\textwidth]{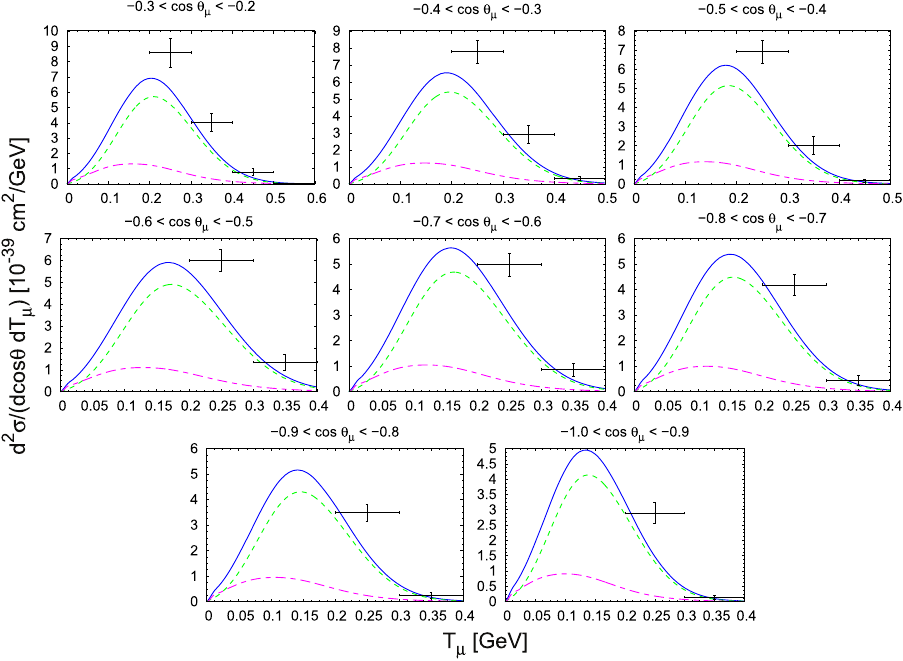}
\caption{(Color online) As for Fig.~\ref{fig:nuMiniBooNE1}, but considering more backward kinematics. The data are from~\cite{PhysRevD.81.092005}.\label{fig:nuMiniBooNE2}}
\end{figure*}

\begin{figure*}[t!]
\centering\includegraphics[width=1.\textwidth]{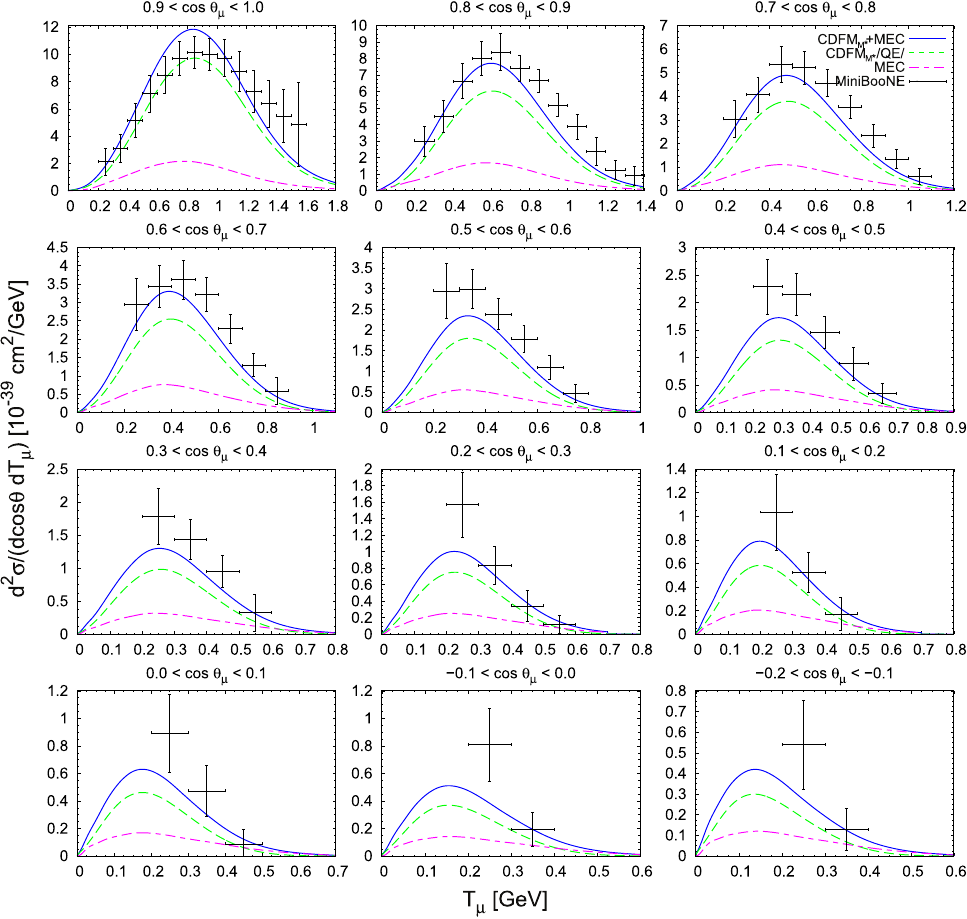}
\caption{(Color online) As for Fig.~\ref{fig:nuMiniBooNE1}, but now for the $\overline{\nu}_\mu$ CCQE process on $^{12}$C. The data are from~\cite{miniboone-ant}.\label{fig:anuMiniBooNE}}
\end{figure*}

\begin{figure*}[t!]
\centering\includegraphics[width=1.\textwidth]{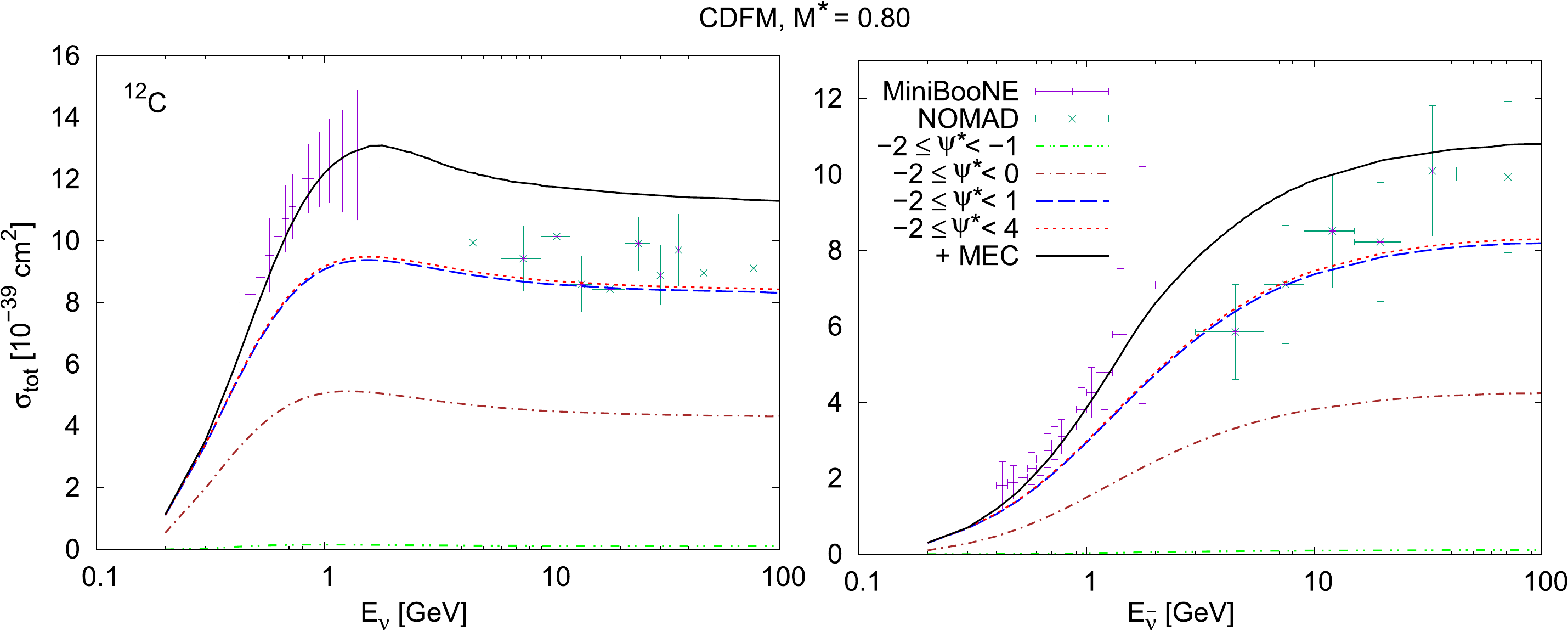}
\caption{(Color online) CCQE $\nu_\mu$--$^{12}$C ($\overline{\nu}_\mu$--$^{12}$C) total cross section per neutron (proton) as a function of the neutrino energy. The left panel (a) corresponds to neutrino cross sections and the right one (b) to antineutrino reactions. The data are from MiniBooNE~\cite{PhysRevD.81.092005, miniboone-ant} and NOMAD~\cite{Lyubushkin:2009} experiments.\label{fig:totCS}}
\end{figure*}

\section{Results and discussions\label{sec:results}}

In this section we use the new scaling function of the CDFM$_{M^*}$ model, given by Eq.~(\ref{eq:8}), to compute lepton scattering cross sections on $^{12}$C. It is important to test CDFM$_{M^*}$ model for inclusive ($e,e'$) scattering before to apply it to neutrino scattering. In Figs.~\ref{fig:CS1}--\ref{fig:CS5} we show the predictions of CDFM$_{M^*}$+MEC contribution (blue solid line) for the ($e, e'$) cross section compared to the experimental data~\cite{RevModPhys.80.189}. Also, the RFG$_{M^*}$+MEC results (green dashed line) are given. The QE contribution within CDFM$_{M^*}$ and RFG$_{M^*}$ are obtained using Eqs.~(\ref{xsecel})--(\ref{GE&GM}) and corresponding scaling functions. The contribution of meson exchange currents (MEC) is presented, separately. The evaluation of the 2p-2h pionic MEC contributions is performed within the RFG model in which a fully Lorentz covariant calculation of the MEC can be performed (see~\cite{DEPACE2003303, PhysRevD.90.033012, Megias:2014qva}). The CDFM$_{M^*}$ model description is quite acceptable using just one free parameter, namely the effective mass $M^*$, which is fixed to $0.8$ in all performed calculations.

In Figs.~\ref{fig:nuMiniBooNE1}--\ref{fig:anuMiniBooNE} we show the double differential cross section averaged over the neutrino and antineutrino energy flux against the kinetic energy of the final muon. The data are taken from the MiniBooNE Collaboration~\cite{PhysRevD.81.092005, miniboone-ant}. We represent a large variety of kinematical situations where each panel refers to results averaged over a particular muon angular bin.

In this work we make use of the 2p-2h MEC model developed in Ref.~\cite{Simo:2016ikv}, which is an extension to the weak sector of the seminal papers~\cite{VanOrden:1980tg, DEPACE2003303, Amaro:2010iu} for the electromagnetic case. The calculation is entirely based on the RFG model and it incorporates the explicit evaluation of the five response functions involved in inclusive neutrino scattering. We use a general parametrization of the MEC responses that significantly reduces the computational time. Its functional form for the cases of $^{12}$C and $^{16}$O is given in Refs.~\cite{PhysRevD.94.093004, PhysRevD.94.013012, Megias2018}.

The results including both QE (obtained within the CDFM$_{M^*}$ model) and 2p--2h MEC are compared with the data in Figs.~\ref{fig:nuMiniBooNE1}--\ref{fig:anuMiniBooNE}. The QE and 2p--2h MEC contributions are presented separately also in the figures. It should be noted the important role played by 2p-2h MEC to describe correctly the experimental data of the order of $\sim$20--25\% of the total response at the maximum. In the neutrino case (Figs.~\ref{fig:nuMiniBooNE1} and~\ref{fig:nuMiniBooNE2}) this relative strength is almost independent of the scattering angle. In the antineutrino case (Fig.~\ref{fig:anuMiniBooNE}) the 2p-2h relative strength gets larger for backward scattering angles. This is due to the fact that the antineutrino cross section involves a destructive interference between the $T$ and $T^\prime$ channels [see Eq.~(\ref{nuxsec})] and is therefore more sensitive to nuclear effects.

Theoretical predictions within the CDFM$_{M^*}$ model including both QE and 2p-2h MEC contributions are in good agreement with the data in most of the kinematical situations explored. Only at scattering angles approaching $90^\circ$ and above one can see a hint of a difference, although in these situations only a small number of data points with large uncertainties exist.

The CDFM$_{M^*}$ results for the total flux-unfolded integrated cross sections per nucleon are given in Fig.~\ref{fig:totCS} being compared with the MiniBooNE~\cite{PhysRevD.81.092005, miniboone-ant} and NOMAD~\cite{Lyubushkin:2009} data (up to $100$~GeV). As can be seen in Fig~\ref{fig:totCS}, the 2p-2h MEC contributions are needed in order to reproduce the MiniBooNE data. Also, the contributions of different parts of the scaling to the total cross sections are presented in Fig.~\ref{fig:totCS}. The main contribution to the cross sections comes from the part of the CDFM$_{M^*}$ scaling function between $-1\leq \psi^* \leq 1$. The CDFM$_{M^*}$ model with 2p--2h MEC clearly overpredicts the NOMAD data. On the contrary, the results without MEC contributions (the pure QE results obtained within CDFM$_{M^*}$ model) are in good agreement with the NOMAD data. This result is consistent with the setup of the NOMAD experiment that, unlike MiniBooNE, can select true QE, rather than the ``QE-like'' events. The role of the 2p-2h MEC is very important at all neutrino energies, getting an almost constant value of the order of $\sim$30$\%-35\%$ compared with the pure QE contribution. Here, we would like to mention that the quasielastic data themselves have been measured not directly but have been deduced from the so-called quasielastic-like data by subtracting a background of events in which pions are firstly produced, but then reabsorbed again. This background was determined from calculations with an event generator. Thus, the final QE + 2p-2h data invariably contain some model dependence~\cite{PhysRevC.87.014602}.

\section{Conclusions\label{sec:conclusions}}

In this paper we have investigated and developed a new scaling approach CDFM$_{M^*}$ using the scaling function derived in the CDFM, the latter being based on the scaling function of the relativistic Fermi gas [Eq.~(\ref{rfg*})]. We use also a new scaling variable $\psi^*$ extracted from the scaling properties of the RMF model in nuclear matter~\cite{Ama15}. Within this model we have obtained a scaling function $f^\text{QE}(\psi^*)$ [Eq.~(\ref{eq:8})] in the QE-region using the empirical density distribution of protons to determine the weight function $|F(x)|^{2}$. The Fermi momentum $k_{F}$ in the CDFM$_{M^*}$ model is not a free parameter and can be obtained by Eq.~(\ref{eq:6}). With the scaling function $f^\text{QE}(\psi^*)$ we calculated the longitudinal and transverse response functions in both ways: using the scaling function from the CDFM$_{M^*}$ and also with the conventional scaling function of the CDFM model (CDFM$_{M^*}$ model with ${M^*=1}$). The CDFM$_{M^*}$ model shows the enhancement of the transverse components of the electromagnetic current. This confirms that using the effective nucleon mass reduction of $M^* = 0.8$ leads to the enhancement of the transverse response (the RMF model includes some dynamical relativistic effects like enhancement of the transverse response due to the lower components of the nucleon spinors). We computed the inclusive $(e,e')$ differential quasielastic cross section with scaling function $f^\text{QE}(\psi^*)$, adding the 2p-2h MEC contribution obtained within the RFG model to the QE results. The theoretical evaluations are shown in  Figs.~\ref{fig:nuMiniBooNE1}--\ref{fig:anuMiniBooNE} and are compared to the world $^{12}$C$(e, e')$ data. It is found a reasonable description of the data.

The next step in our study was to use the scaling function $f^\text{QE}(\psi^*)$ obtained in CDFM$_{M^*}$ model to predict the CCQE (anti)neutrino-nucleus scattering processes that are of interest for (anti)neutrino oscillation experiments. These calculations, based on the impulse approximation, are complemented with the contributions given by two-body weak meson exchange currents, giving rise to two-particle two-hole excitations. The model is applied to the MiniBooNE experiment (see Figs.~\ref{fig:nuMiniBooNE1}--\ref{fig:totCS}). We find that the scaling approach CDFM$_{M^*}$ including both QE and 2p-2h MEC leads to results that are in good agreement with the data in most of the kinematical situations explored in this research. One can see the contribution ascribed to the 2p-2h MEC effects that can be even larger than $\sim$30$\%-35\%$ compared with the pure QE responses. This proves without ambiguity the essential role played by 2p-2h MEC in providing a successful description of (anti)neutrino-nucleus scattering data. The results in this work can be considered as a test of the reliability of the present CDFM$_{M^*}$ model. The present study gives us a confidence to extend the use of this model to predict CCQE $\nu$($\overline{\nu}$)-nucleus scattering processes to other nucleus and energies and to other processes, such as the semi-inclusive CC$\nu$ reactions and neutral current processes.

It is crucial to note that the parametrization of the 2p2h MEC employed in this study is derived from the RFG model with an effective mass of $M^*=1$. An alternative parametrization for electroweak 2p2h MEC responses, calculated in the RMF with an effective mass of $M^*=0.8$, has been introduced in Refs.~\cite{PhysRevC.104.025501, PhysRevD.104.113006}. This new parametrization, which employs a semiempirical formula, could provide benefits over the CDFM$_{M^*}$ model. Therefore, the initial step to enhance our model involves adopting this alternative parametrization to evaluate its potential influence on the subsequent research.

It is shown in our work that the CDFM$_{M^*}$ model describes successfully inclusive $(e,e')$ and $\nu$($\overline{\nu}$) CCQE quasielastic cross section on the basis of the new scaling variable $\psi^*$ [Eq.~(\ref{psi*})], of the empirical density distribution of protons to determine the weight function $|F(x)|^{2}$ [Eq.~(\ref{eq:2})], and of the corresponding scaling function $f^\text{QE}(\psi^*)$ [Eq.~(\ref{eq:8})]. We note that in the CDFM$_{M^*}$ model an effective mass $M^*=m_N^*/m_N=0.8$ is used. The latter is originating from the interacting RFG model in which the vector and scalar potentials generate the effective mass of the nucleon in medium. We should emphasize that the CDFM$_{M^*}$ scaling function keeps the gauge invariance (that is not the case in the SuSA approach) and describes the dynamical enhancement of the lower components of the relativistic spinors, as well as the transverse response function. In addition, we note the important fact that in the CDFM$_{M^*}$ model the weight and scaling functions are normalized to unity. It is pointed out that the constructed realistic CDFM$_{M^*}$ scaling function is an essential ingredient in this approach for the description of the processes of lepton scattering from nuclei. Another interesting future project will be to extend the scaling approach using a constructed realistic CDFM$_{M^*}$ scaling function to obtain predictions for the charge-changing neutrino and antineutrino scattering from nuclei in the $\Delta$-region.

\begin{acknowledgements}
This work was partially supported by the Bulgarian National Science Fund under contract No. KP-06-N38/1.
\end{acknowledgements}

\bibliography{biblio}

\end{document}